\documentclass[aps,preprint]{revtex4}%
\usepackage{amsfonts}
\usepackage{amsmath}
\usepackage{amssymb}
\usepackage{graphicx}
\usepackage{natbib}%
\providecommand{\U}[1]{\protect \rule{.1in}{.1in}}

\begin{document}
\preprint{ }

\begin{abstract}
We expand the existing polaron response theory, expressed within the
Mori-Zwanzig projection operator formalism applicable for the transfer of
arbitrary energy and zero momentum, to the the case of finite momentum
exchange. A general formula is then derived which can be used to calculate the
response of a system to a probe that transfers both momentum and energy to the
system. The main extension of the existing polaron response theory is the
finite momentum exchange which was not needed until now\ since it is
negligible for optical absorption. However, this formalism is needed to
calculate the response of the polaronic system consisting of an impurity in a
Bose-Einstein condensate to Bragg spectroscopy. We show that the well-known
features that appear in the optical absorption of the solid state Fr\"{o}hlich
polaron are also present in the Bragg response of the BEC-impurity polaron.
The f-sum rule is written in a form suitable to provide an independent
consistency test for our results. The effect of lifetime broadening on the
BEC-impurity spectrum is examined. The results derived here are discussed in
the framework of an experimental realization consisting of a lithium impurity
in a sodium condensate.

\end{abstract}
\title{Response of the polaron system consisting of an impurity in a Bose-Einstein
condensate to Bragg spectroscopy}
\author{W. Casteels$^{1}$, J. Tempere$^{1,2}$ and J. T. Devreese$^{1}$}
\affiliation{$^{1}$TQC, Universiteit Antwerpen, Groenenborgerlaan 171, B2020 Antwerpen, Belgium}
\affiliation{$^{2}$Lyman Laboratory of Physics, Harvard University, Cambridge,
Massachusetts 02138, USA}
\maketitle

\section{Introduction}

The experimental realization of impurities in a Bose-Einstein condensate (BEC)
\cite{CavityNature, PhysRevLett.85.483, 063118} and the possibility to produce
quantum degenerate atomic mixtures \cite{PhysRevLett.95.170408,
PhysRevLett.89.190404, PhysRevLett.88.160401}\ has resulted in an interest in
the physics of impurities in a quantum gas. This has led to an investigation
of the change of the properties of the bare impurities as a result of the
interactions with the condensate (such as the effective interaction
\cite{PhysRevA.72.023616, PhysRevA.61.053601} and the effective mass
\cite{PhysRevA.70.013608, girardeau:279, Gross1962234}) and a study of the
self-trapping of the impurity \cite{PhysRevA.73.043608, PhysRevB.46.301}. Also
the system of a spin down fermion in a sea of spin up fermions, usually
indicated as the Fermi polaron, has been investigated which resulted in a good
agreement between theory \cite{PhysRevA.80.033607, PhysRevB.77.020408} and
experiments \cite{PhysRevLett.103.170402, PhysRevLett.102.230402}. Furthermore
it was shown that when the Bogoliubov approximation is valid the system of an
impurity in a BEC can be mapped to the Fr\"{o}hlich polaron system
\cite{PhysRevLett.96.210401, PhysRevA.73.063604, PhysRevB.80.184504} which
shows that this system can be added to the list of condensed matter systems
that can be imitated in the context of quantum gases \cite{RevModPhys.80.885}.
The experimental realization of quantum degenerate atomic mixtures in an
optical lattice \cite{PhysRevLett.102.030408, PhysRevLett.96.180403,
PhysRevLett.96.180402} has led to the theoretical study of the polaronic
properties of impurities in a BEC where the influence of the optical lattice
is felt only by the impurities \cite{PhysRevA.76.011605,
1367-2630-10-3-033015} or by both the impurities and the condensate
\cite{2010arXiv1009.0675P}.

In the context of solid state physics the Fr\"{o}hlich polaron consists of a
charge carier (electron, hole) interacting with the phonons in an ionic
crystal or a polar semiconductor \cite{LandauPekar, Frolich3}. This
Fr\"{o}hlich polaron Hamiltonian can not be diagonalised exactly and several
approximation methods have been developed for it. The most general way to
study the system is through a variational principle within the path integral
formalism which was developed by Feynman \cite{PhysRev.97.660}.\ The internal
excitation spectrum of the Fr\"{o}hlich polaron was revealed through the
calculation of the optical absorption by Devreese \textit{et al}.
\cite{PhysRevB.5.2367}, based on the Feynman path integral method and a
response formalism introduced in Ref. \cite{PhysRev.127.1004}. Recently the
excitation spectrum of the Fr\"{o}hlich polaron Hamiltonian was also studied
numerically with Diagrammatic Quantum Monte-Carlo numerical techniques and
these results showed deviations in the large coupling optical absorption
linewidth with the theoretical results of \cite{PhysRevB.5.2367}, see Ref.
\cite{PhysRevLett.91.236401}. At weak and intermediate coupling the analytical
theory of Ref. \cite{PhysRevB.5.2367} agrees well with the numerical
simulations of Ref. \cite{PhysRevLett.91.236401} and this optical absorption
spectrum was also studied experimentally \cite{PhysRevB.64.104504}. So far the
strong coupling regime could not be probed experimentally since the largest
attainable Fr\"{o}hlich polaron coupling constant in any solid is not large
enough. The large coupling behavior remains an important question since a
better understanding of the intermediate and strong coupling regimes might be
useful to elucidate the role of polarons and bipolarons in unconventional
pairing mechanisms e.g. for high-temperature superconductivity
\cite{PhysRevB.77.094502}. Furthermore it was shown that at intermediate
coupling the introduction of an \textit{ad hoc} lifetime $\tau$ for the
eigenstates of the polaron model system\ is an improvement of the theory which
is called the extended memory function formalism \cite{PhysRevLett.96.136405}.

If the Bogoliubov approximation is applicable the Hamiltonian of an impurity
in a BEC can be written as the sum of the mean field energy of the homogeneous
condensate and the Fr\"{o}hlich polaron Hamiltonian that describes the
fluctuations \cite{PhysRevB.80.184504}:%
\begin{equation}
\widehat{H}_{pol}=\frac{\widehat{p}^{2}}{2m_{I}}+\sum_{\vec{k}\neq0}%
\hbar \omega_{\vec{k}}\widehat{b}_{\vec{k}}^{\dag}\widehat{b}_{\vec{k}}%
+\sum_{\vec{k}\neq0}V_{\vec{k}}e^{i\vec{k}.\widehat{r}}\left(  \widehat
{b}_{\vec{k}}+\widehat{b}_{-\vec{k}}^{\dag}\right)  ,
\end{equation}
which describes the interaction between the impurity and the Bogoliubov
excitations where $\widehat{r}$ and $\widehat{p}$ represent the position and
momentum of the impurity with mass $m_{I}$, $\widehat{b}_{\vec{k}}$ and
$\widehat{b}_{\vec{k}}^{\dag}$\ are the annihilation and creation operators
for the Bogoliubov excitations,\ $\omega_{\vec{k}}$ is the Bogoliubov
frequency:%
\begin{equation}
\omega_{\vec{k}}=ck\sqrt{1+\left(  \xi k\right)  ^{2}/2},
\end{equation}
and $V_{\vec{k}}$ is the interaction amplitude:%
\begin{equation}
V_{\vec{k}}=\sqrt{N_{0}}\left[  \frac{\left(  \xi k\right)  ^{2}}{\left(  \xi
k\right)  ^{2}+2}\right]  ^{1/4}g_{IB}.
\end{equation}
In the above expressions use was made of the definition of the healing length
of the condensate: $\xi=1/\sqrt{8\pi a_{BB}n_{0}}$ with $n_{0}=N_{0}/V$ the
condensate density and $a_{BB}$ the boson-boson scattering length. We also
used the expression for the speed of sound in the condensate: $c=\hbar/\left(
\sqrt{2}m_{B}\xi \right)  $. In \cite{PhysRevB.80.184504} the Feynman
variational path integral technique was applied to this Hamiltonian and an
upper bound for the free energy was calculated by using the Jensen-Feynman
inequality with the Feynman model system. This model system consists of the
impurity mass coupled to another mass $M$ by a spring with frequency $\Omega$.
The parameters $M$ and $\Omega$ are then used to minimize the free energy. It
followed that the Fr\"{o}hlich polaron coupling parameter in this case is
given by:%
\begin{equation}
\alpha=\frac{a_{IB}^{2}}{a_{BB}\xi},
\end{equation}
with $a_{IB}$ the impurity-boson scattering length. Together with the ratio
between the masses of the impurity and the bosons this coupling parameter
fully determines the static properties of the specific BEC-impurity polaron
system. Depending on the value of this coupling strength two regimes were
identified: a strong coupling regime which has properties that suggested a
polaronic self-trapped state and a weak coupling regime which suggested a free
polaron. Since this coupling parameter depends strongly on the scattering
lengths, which can be tuned externally by a magnetic field through a Feshbach
resonance (see eg. \cite{Pitaevskii}), it may be possible to tune the system
experimentally to the regime of strong coupling. This technique might enable
the experimental realization of the strong coupling regime and reveal the
internal structure of the Fr\"{o}hlich polaron Hamiltonian.

\begin{figure}
[ptb]
\begin{center}
\includegraphics[
height=4.3691cm,
width=6.0934cm
]%
{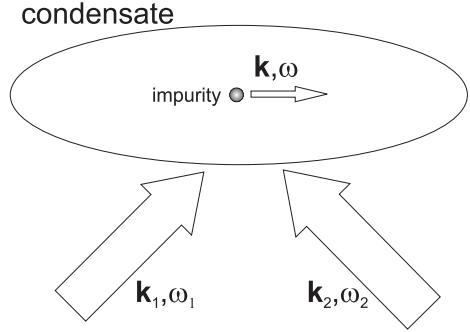}%
\caption{Schematical picture of a typical experimental set-up for Bragg
spectroscopy. Two lasers with momentum $\vec{k}_{1}$ and $\vec{k}_{2}$ and
energy $\hbar \omega_{1}$ and $\hbar \omega_{2}$ are impinged upon the impurity
in the condensate which gains an energy $\hbar \omega=\hbar \omega_{1}%
-\hbar \omega_{2}$ and a momentum $\vec{k}=\vec{k}_{1}-\vec{k}_{2}.$}%
\label{fig: Bragg}%
\end{center}
\end{figure}

Since, as opposed to the original Fr\"{o}hlich polaron in solid state, the
BEC-impurity polaron system is not charged it is not helpful to perform
optical absorption measurements on this system. The method to probe the system
is addressed in the present paper and will be shown to be Bragg spectroscopy.
This has proven to be a very succesfull tool to probe the structure of BEC's
(see eg. \cite{PhysRevLett.83.2876} and \cite{PhysRevLett.88.120407}). It is
realized by probing the system with two laser beams with wave vectors $\vec
{k}_{1}$ and $\vec{k}_{2}$ and frequencies $\omega_{1}$ and $\omega_{2}$. An
atom can then undergo a stimulated scattering event by absorbing a photon from
beam 1 and emitting it in beam 2 which changes its momentum by $\vec{k}%
=\vec{k}_{1}-\vec{k}_{2}$ and its energy by $\hbar \omega=\hbar \omega_{1}%
-\hbar \omega_{2}$. See figure \ref{fig: Bragg} for a schematical picture of a
typical experimental set-up. A possible way to measure the response of the
system is by performing a time-of-flight experiment and count the number of
atoms $N_{Bragg}$ that have gained a momentum $\vec{k}$. Within the formalism
of linear response theory this number is given by \cite{Pitaevskii}:%
\begin{equation}
N_{Bragg}=\frac{2}{\hbar}\left(  \frac{V}{2}\right)  ^{2}\tau \operatorname{Im}%
\chi \left(  \omega,\vec{k}\right)  ,\label{ScatAtoms}%
\end{equation}%
with $\tau$ the duration of the pulse, $V$ the amplitude of the laser induced
potential and $\chi \left(  \omega,\vec{k}\right)  $ the density response
function, defined by:%
\begin{equation}
\chi \left(  \omega,\vec{k}\right)  =-\frac{1}{\hbar}\mathcal{Z}^{-1}\sum
_{m,n}e^{-\beta E_{m}}\left[  \frac{\left \vert \left \langle m\left \vert
\rho_{\vec{k}}\right \vert n\right \rangle \right \vert ^{2}}{\omega-\omega
_{nm}+i\delta}-\frac{\left \vert \left \langle m\left \vert \rho_{\vec{k}}^{\dag
}\right \vert n\right \rangle \right \vert ^{2}}{\omega+\omega_{nm}+i\delta
}\right]  ,
\end{equation}
with $\rho_{\vec{k}}$ the induced density of the impurity, $\mathcal{Z}$ the
partition function, $|n>$ the eigenstates of the unperturbed system and
$\omega_{nm}=\left(  E_{n}-E_{m}\right)  /\hbar$ the transition frequencies.
This density response function can be rewritten as the Fourier transform of
the retarded density-density correlation function:%
\begin{equation}
\chi \left(  \omega,\vec{k}\right)  =\frac{i}{\hbar}\int_{0}^{\infty
}dte^{i\omega t}\left \langle \left[  \rho_{\vec{k}}\left(  t\right)
,\rho_{\vec{k}}^{\dag}\right]  \right \rangle .\label{dens-dens}%
\end{equation}
The retarded density-density correlation function is needed to obtain the
response of a system to Bragg spectroscopy.

We will start by using the Mori formalism to calculate a general expression
for the density-density correlation function (\ref{dens-dens}) and then apply
it to the generic polaron system with arbitrary dispersion and interaction
amplitude. Then we show that if we calculate from this expression the
current-current correlation for the $\vec{k}=0$ limit we find the known result
for the optical absorption of Ref. \cite{PhysRevB.5.2367}. Next we introduce
the dispersion and interaction amplitude for the BEC-impurity polaron and
obtain results for the response to Bragg spectroscopy. The result is expressed
as a double integral over highly oscillatory integrands that has to be
calculated numerically and which can give numerical problems. To overcome
these problems we will then rewrite our expressions in another representation
which is suitable for numerical work. We then write down the f-sum rule to
check our results and find that at low temperature there is a relatively large
recoil contribution for $\omega \rightarrow0$ which numerically has to be added
separately in the sum rule. This behavior was also found in the optical
absorption of the solid state Fr\"{o}hlich polaron and in that case there is a
relation with the effective mass of the polaron \cite{PhysRevB.15.1212}.
Recently this sum rule was used to experimentally determine the effective mass
of the solid state Fr\"{o}hlich polaron, see Refs.
\cite{PhysRevLett.100.226403} and \cite{PhysRevB.81.125119}. Next, following
De Filippis \textit{et al }\cite{PhysRevLett.96.136405}, we introduce a
lifetime $\tau$ and examine the dependence of the theory on this lifetime. The
results are analyzed for the experimental setup consisting of a Lithium
impurity in a sodium condensate which is the same system that was examined in
Ref. \cite{PhysRevB.80.184504}.

\section{Application of the Mori formalism to the density-density correlation}

The Mori-Zwanzig projection operator technique (see \cite{PTP.33.423} and
\cite{PhysRev.124.983}) is a well established technique to calculate
correlation functions and has been applied before to the Fr\"{o}hlich polaron
Hamiltonian to calculate the optical absorption by Peeters and Devreese
\cite{PhysRevB.28.6051} (for a detailed description we refer to
\cite{ZomerschoolDevreese}) . This technique can also be applied to the
calculation of the density-density correlation function. There is however one
obstacle that we encountered when defining the Mori projection operator. The
reason is that the common definition of this operator is ill-defined since it
leads to a vanishing density-density commutator. This problem was already
noticed by Ichiyanagi \cite{JPSJ.32.604} and he suggested the following
definition for the action of the Mori projection operator $P$ on an arbitrary
operator $A$:%
\begin{equation}
PA=\frac{\left \langle \left[  A,\dot{\rho}_{\vec{k}}^{\dag}\right]
\right \rangle }{\left \langle \left[  \rho_{\vec{k}},\dot{\rho}_{\vec{k}}%
^{\dag}\right]  \right \rangle }\rho_{\vec{k}}.
\end{equation}
In our calculations we have followed this suggestion and found that the
density response function can be written as:%
\begin{equation}
\chi \left(  \omega,\vec{k}\right)  =\frac{i}{\hbar}\frac{\varphi}{\omega
^{2}+O-\Sigma \left(  \omega,\vec{k}\right)  }, \label{Mori result}%
\end{equation}
with:%
\begin{align}
\varphi &  =\left \langle \left[  \rho_{\vec{k}},\dot{\rho}_{\vec{k}}^{\dag
}\right]  \right \rangle ;\\
O  &  =\frac{\omega}{\varphi}\left \langle \left[  L\rho_{\vec{k}},\dot{\rho
}_{\vec{k}}^{\dag}\right]  \right \rangle ;\\
\Sigma \left(  \omega,\vec{k}\right)   &  =-\frac{1}{\varphi}\int_{0}^{\infty
}dt\left(  1-e^{i\omega t}\right)  \left \langle \left[  \mathcal{B}_{\vec{k}%
}\left(  t\right)  ,\mathcal{B}_{\vec{k}}^{\dag}\right]  \right \rangle .
\end{align}
Where $L$ is the Liouville operator and $\Sigma \left(  \omega,\vec{k}\right)
$ is known as the memory function or self energy. Furthermore we also
introduced the operator $\mathcal{B}_{\vec{k}}\left(  t\right)  $ defined as:%
\begin{equation}
\mathcal{B}_{\vec{k}}=QL\dot{\rho}_{\vec{k}},
\end{equation}
with $Q$ the complementary Mori projection operator: $Q=1-P$. The time
dependence of $\mathcal{B}_{\vec{k}}\left(  t\right)  $ is governed by a new
Liouville operator defined by $\mathcal{L=}QLQ$. It is important to note here
that until this point no approximations were made and the result
(\ref{Mori result}) for the linear density response function is exact. Also,
to derive the formula (\ref{Mori result}) we have not introduced a specific
system and so it is a general result for the density response function and can
also be applied to calculate the response of other systems to probes with an
arbitrary energy and momentum exchange. For example the final state of neutron
scattering is also determined by the density reponse function
\cite{PhysRev.95.249}. We now consider the Fr\"{o}hlich polaron Hamiltonian
and follow an approach similar to that used in Ref. \cite{PhysRevB.28.6051}.
This consists in replacing the new Liouville operator $\mathcal{L}$ by the sum
of the Liouville operator of free Bogoliubov excitations $L_{Bog}$ and the
Liouville operator of the Feynman modelsystem $L_{F}$: $\mathcal{L\rightarrow
}L_{Bog}+L_{F}$. Within this approximation we can calculate the different
quantities in the expression (\ref{Mori result}) which gives $O=0$ and:%
\begin{equation}
\varphi=i\frac{\hbar k^{2}}{m_{I}}N,
\end{equation}
where $N$ is the number of impurities. The memory function is then given by:%
\begin{align}
\Sigma \left(  \omega,\vec{k}\right)   &  =\frac{2}{m_{I}N\hbar}\sum_{\vec
{q}\neq0}\left \vert V_{\vec{q}}\right \vert ^{2}\frac{\left(  \vec{k}.\vec
{q}\right)  ^{2}}{k^{2}}\int_{0}^{\infty}dt\left(  1-e^{i\omega t}\right)
\nonumber \\
&  \times \operatorname{Im}\left \{  \left[  e^{i\omega_{\vec{q}}t}+2\cos \left(
\omega_{\vec{q}}t\right)  n_{\vec{q}}\right]  \exp \left[  -\left(  \vec
{k}+\vec{q}\right)  ^{2}D\left(  t\right)  \right]  \right \}  ,
\end{align}
with $n_{\vec{q}}$ the Bose-Einstein distrubution, i.e. the number of
Bogoliubov excitations with frequency $\omega_{\vec{q}}$ and $D\left(
t\right)  $ describes the Feynman model system and is given by:%
\begin{align}
D\left(  t\right)   &  =\frac{t^{2}}{2\beta \left(  m+M\right)  }-i\frac{\hbar
}{2\left(  m+M\right)  }t\nonumber \\
&  +\frac{\hbar M}{2m\Omega(m+M)}\left(  1-\exp \left(  i\Omega t\right)
+4\sin^{2}\left(  \frac{\Omega t}{2}\right)  n\left(  \Omega \right)  \right)
, \label{Dfun}%
\end{align}
with $M$ and $\Omega$ the variational parameters. The values of these
parameters are deduced from the minimalisation of the free energy
\cite{PhysRevB.80.184504}.

\section{Link with the optical absorption of the solid state Fr\"{o}hlich
polaron}

A well-known result from linear response theory and the Kubo formalism is that
the optical absorption of a system can be expressed as a current-current
correlation function \cite{Mahan}. Since the current and the density are
related to each other through the continuity equation it is possible to
calculate the current-current correlation function from the density-density
correlation function. For the process of optical absorption the photon
momentum is negligible in comparison to the other momenta involved in the
polaron problem, which is not the case for Bragg spectroscopy. To obtain the
optical absorption we have to take the limit $\vec{k}\rightarrow0$ which gives
for the memory function:%
\begin{align}
\Sigma \left(  \omega,0\right)   &  =\frac{2}{3m_{I}N\hbar}\sum_{\vec{q}\neq
0}\left \vert V_{\vec{q}}\right \vert ^{2}q^{2}\int_{0}^{\infty}dt\left(
1-e^{i\omega t}\right) \nonumber \\
&  \times \operatorname{Im}\left \{  \left[  e^{i\omega_{\vec{q}}t}+2\cos \left(
\omega_{\vec{q}}t\right)  n_{\vec{q}}\right]  \exp \left[  -q^{2}D\left(
t\right)  \right]  \right \}  .
\end{align}
This is just the memory function $\Sigma \left(  \omega \right)  $ calculated in
Ref. \cite{PhysRevB.28.6051} multiplied by $\omega$:%
\begin{equation}
\Sigma \left(  \omega,0\right)  =\omega \Sigma \left(  \omega \right)  .
\end{equation}
For the solid state Fr\"{o}hlich polaron it is the current-current correlation
that is needed for the optical absorption. By applying two partial
integrations and making use of the continuity equation one can find a simple
relation between the two:%
\begin{equation}
\chi \left(  \omega,\vec{k}\right)  =-\frac{i}{\hbar \omega^{2}}\int_{0}%
^{\infty}dte^{i\omega t}\left \langle \left[  \vec{k}.\vec{j}_{\vec{k}}\left(
t\right)  ,\vec{k}.\vec{j}_{\vec{k}}^{\dag}\right]  \right \rangle
\end{equation}
We now make use of the standard Kubo formula for the optical conductivity (see
e.g. \cite{Mahan}):%
\begin{align}
\operatorname{Re}\left[  \sigma \left(  \omega \right)  \right]   &
=\operatorname{Re}\left[  i\frac{e^{2}}{Vm\omega}+\frac{1}{\hbar \omega}%
\int_{0}^{\infty}dte^{i\omega t}\left \langle \left[  j_{x}\left(  t\right)
,j_{x}\right]  \right \rangle \right] \nonumber \\
&  =\lim_{\vec{k}\rightarrow0}\operatorname{Im}\left[  -\frac{\hbar \omega
}{k^{2}}\chi \left(  \omega,\vec{k}\right)  \right] \nonumber \\
&  =\operatorname{Im}\left[  -\frac{1}{m_{I}}\frac{1}{\omega-\Sigma \left(
\omega \right)  }\right]  ,
\end{align}
this is the result that was found in Ref. \cite{PhysRevB.5.2367}\ from the
Feynman path integral formalism and rederived in Ref. \cite{PhysRevB.28.6051}
using Mori's formalism.

\section{Results for the BEC-impurity polaron}

In this section we will introduce the dispersion and the interaction amplitude
for the BEC-impurity system and obtain an expression for the response of the
system to a Bragg pulse. From expression (\ref{ScatAtoms}) it follows that the
response of the system is characterised by the imaginary part of the density
response function which can be written as:%
\begin{equation}
\operatorname{Im}\left[  \chi \left(  \omega,\vec{k}\right)  \right]
=-\frac{k^{2}}{m_{I}}N\frac{\operatorname{Im}\left[  \Sigma \left(  \omega
,\vec{k}\right)  \right]  }{\left(  \omega^{2}-\operatorname{Re}\left[
\Sigma \left(  \omega,\vec{k}\right)  \right]  \right)  ^{2}+\left(
\operatorname{Im}\left[  \Sigma \left(  \omega,\vec{k}\right)  \right]
\right)  ^{2}}. \label{repons}%
\end{equation}
This particular form for the response function is general for the Fr\"{o}hlich
polaron and was first introduced in \cite{PhysRevB.5.2367}. The memory
function for the BEC-impurity system is given by (with polaronic units i.e.
$\hbar=m_{I}=\xi=1$):%
\begin{align}
\Sigma \left(  \omega,\vec{k}\right)   &  =\frac{1}{N}\frac{\alpha}{4\pi
}\left(  \frac{m_{B}+1}{m_{B}}\right)  ^{2}\int_{0}^{\infty}dq\frac{q^{5}%
}{\sqrt{q+2}}\int_{0}^{\infty}d\widetilde{t}\left(  1-e^{i\omega t}\right)
\nonumber \\
&  \times \operatorname{Im}\left[  \frac{\left[  e^{i\omega_{\vec{q}}t}%
+2\cos \left(  \omega_{\vec{q}}t\right)  n_{\vec{q}}\right]  }{\left[
2kqD\left(  t\right)  \right]  ^{3}}\left(  \exp \left[  -\left(  k-q\right)
^{2}D\left(  t\right)  \right]  \left \{  \left[  1-2kqD\left(  t\right)
\right]  ^{2}+1\right \}  \right.  \right. \nonumber \\
&  \left.  \left.  -\exp \left[  -\left(  k+q\right)  ^{2}D\left(  t\right)
\right]  \left \{  \left[  1+2kqD\left(  t\right)  \right]  ^{2}+1\right \}
\right)  \right]  . \label{SelfEnergie}%
\end{align}
The double integral in expression (\ref{SelfEnergie}) could not be calculated
analytically and it was computed numerically.

\section{Second representation for the imaginary part of the memory function}

Due to the highly oscillatory behavior of the integrand it turns out that
expression (\ref{SelfEnergie}) is not of a form efficient for numerical
calculation of the imaginary part of the memory function. Therefore we develop
an alternative representation for the memory function which allows to perform
the time integration analytically. This provides a representation for the
imaginary part that is well-suited for numerical calculations. The real part
of the memory function turns out to be far more involved but it seems that in
this case we can use expression (\ref{SelfEnergie}). We will follow a similar
approach as was used in Ref. \cite{PhysRevB.5.2367} and also in Ref.
\cite{PhysRevB.28.6051}. We start by rewriting (\ref{Dfun}) as:%
\begin{align}
D\left(  t\right)   &  =\frac{t^{2}}{2\beta \left(  m+M\right)  }-i\frac{\hbar
}{2\left(  m+M\right)  }t\nonumber \\
&  +\frac{\hbar M}{2m\Omega(m+M)}\left[  \coth \left[  \frac{\hbar \beta \Omega
}{2}\right]  -\left(  1+n\left(  \Omega \right)  \right)  \exp \left(  i\Omega
t\right)  -n\left(  \Omega \right)  \exp \left(  -i\Omega t\right)  \right]  .
\end{align}
After two Taylor expansions for the two exponentials and writing the term with
$t^{2}$ as a Gaussian integral we obtain:%
\begin{align}
\exp \left(  -\left(  \vec{k}+\vec{q}\right)  ^{2}D\left(  t\right)  \right)
&  =\left(  \frac{\beta}{2\pi \left(  m+M\right)  }\right)  ^{1/2}e^{-\left(
\vec{k}+\vec{q}\right)  ^{2}\frac{\hbar M}{2m\Omega(m+M)}\coth \left[
\frac{\hbar \beta \Omega}{2}\right]  }\nonumber \\
&  \times \sum_{n,n^{\prime}}^{\infty}\frac{1}{n!}\frac{1}{n^{\prime}!}\left[
\frac{\hbar M}{2m\Omega(m+M)}\left(  1+n\left(  \Omega \right)  \right)
\right]  ^{n}\left[  \frac{\hbar M}{2m\Omega(m+M)}n\left(  \Omega \right)
\right]  ^{n^{\prime}}\nonumber \\
&  \times \left(  \vec{k}+\vec{q}\right)  ^{2\left(  n+n^{\prime}\right)  }%
\int_{-\infty}^{\infty}dp\exp \left \{  -\frac{\beta p^{2}}{2\left(  M+m\right)
}\right \} \nonumber \\
&  \times \exp \left \{  it\left[  \frac{\hbar \left(  \vec{k}+\vec{q}\right)
^{2}}{2\left(  m+M\right)  }\pm \frac{p\left \vert \vec{k}+\vec{q}\right \vert
}{m+M}+\left(  n-n^{\prime}\right)  \Omega \right]  \right \}  .
\end{align}
Furthermore the memory function can be written as:%
\begin{align}
\Sigma \left(  \omega,\vec{k}\right)   &  =\frac{\hbar \omega}{\chi}\int
_{0}^{\infty}dte^{i\omega t}\int_{0}^{\beta}d\lambda \left \langle
\mathcal{B}_{\vec{k}}\left(  t-i\hbar \lambda \right)  \mathcal{B}_{\vec{k}%
}^{\dag}\right \rangle \nonumber \\
&  =-\frac{i\omega}{Nm}\sum_{\vec{q}\neq0}\left \vert V_{\vec{q}}\right \vert
^{2}\frac{\left(  \vec{k}.\vec{q}\right)  ^{2}}{k^{2}}\int_{0}^{\infty
}dte^{i\omega t}\int_{0}^{\beta}d\lambda \left[  \left(  1+n_{\vec{q}}\right)
e^{-i\omega_{\vec{q}}\left(  t-i\hbar \lambda \right)  }+n_{\vec{q}}%
e^{i\omega_{\vec{q}}\left(  t-i\hbar \lambda \right)  }\right] \nonumber \\
&  \times \exp \left[  -\left(  \vec{k}+\vec{q}\right)  ^{2}D\left(
-t+i\hbar \lambda \right)  \right]  .
\end{align}
Now the time integration is straightforward:%
\begin{align}
\Sigma \left(  \omega,\vec{k}\right)   &  =\frac{\omega}{Nm}\left(  \frac
{\beta}{2\pi \left(  m+M\right)  }\right)  ^{1/2}\sum_{\vec{q}\neq0}\left \vert
V_{\vec{q}}\right \vert ^{2}e^{-\left(  \vec{k}+\vec{q}\right)  ^{2}%
a^{2}\left(  \beta \right)  }\frac{\left(  \vec{k}.\vec{q}\right)  ^{2}}{k^{2}%
}\sum_{n,n^{\prime}}^{\infty}B\left(  \beta,n,n^{\prime}\right) \nonumber \\
&  \times \left(  \vec{k}+\vec{q}\right)  ^{2\left(  n+n^{\prime}\right)  }%
\int_{-\infty}^{\infty}dpe^{-\frac{\beta p^{2}}{2\left(  M+m\right)  }}\left[
\frac{1-e^{-\hbar \beta \nu_{-}}}{\hbar \nu_{-}}\frac{n_{\vec{q}}}{\omega-\nu
_{-}+i\varepsilon}+\frac{1-e^{-\hbar \beta \nu_{+}}}{\hbar \nu_{+}}%
\frac{1+n_{\vec{q}}}{\omega-\nu_{+}+i\varepsilon}\right]  ,
\label{Self2deRepr}%
\end{align}
with the following notation:%
\begin{align}
a^{2}\left(  \beta \right)   &  =\frac{\hbar M}{2m\Omega(m+M)}\coth \left[
\frac{\hbar \beta \Omega}{2}\right]  ;\\
B\left(  \beta,n,n^{\prime}\right)   &  =\frac{1}{n!}\frac{1}{n^{\prime}%
!}\left[  a^{2}\left(  1+n\left(  \Omega \right)  \right)  \right]  ^{n}\left[
a^{2}n\left(  \Omega \right)  \right]  ^{n^{\prime}};\\
\nu_{\pm}  &  =\pm \omega_{\vec{q}}+\frac{\hbar \left(  \vec{k}+\vec{q}\right)
^{2}}{2\left(  m+M\right)  }\pm \frac{p\left \vert \vec{k}+\vec{q}\right \vert
}{m+M}+\left(  n-n^{\prime}\right)  \Omega.
\end{align}
The memory function is now split in an imaginary and a real part with the
formula of Plemelj. The integral over $p$ in the imaginary part of
(\ref{Self2deRepr}) can be performed easily by using the delta function.
Taking into account the dispersion and the interaction amplitude, this yields
(with polaronic units):%
\begin{align}
\operatorname{Im}\left[  \Sigma \left(  \omega,\vec{k}\right)  \right]   &
=-\frac{1}{N}\frac{\alpha}{8}\left(  \frac{m_{B}+1}{m_{B}}\right)  ^{2}\left(
\frac{\beta \left(  1+M\right)  }{2\pi}\right)  ^{1/2}\nonumber \\
&  \times \int_{0}^{\infty}dqq^{2}\int_{-1}^{1}dx\sqrt{\frac{q^{2}}{q^{2}+2}%
}e^{-\left(  k^{2}+q^{2}+2qkx\right)  a^{2}\left(  \beta \right)  }\left(
qx\right)  ^{2}\nonumber \\
&  \times \sum_{n,n^{\prime}}^{\infty}B\left(  \beta,n,n^{\prime}\right)
\left(  k^{2}+q^{2}+2qkx\right)  ^{\left(  n+n^{\prime}\right)  -1/2}\left(
1-e^{-\beta \omega}\right) \nonumber \\
&  \times \left[  n_{\vec{q}}\exp \left \{  -\frac{\beta \left(  M+1\right)
A_{n}^{-}\left(  \omega \right)  ^{2}}{2\left(  k^{2}+q^{2}+2qkx\right)
}\right \}  +\left(  1+n_{\vec{q}}\right)  \exp \left \{  -\frac{\beta \left(
M+1\right)  A_{n}^{+}\left(  \omega \right)  ^{2}}{2\left(  k^{2}%
+q^{2}+2qkx\right)  }\right \}  \right]  , \label{TweedeRepr}%
\end{align}
with:%
\[
A_{n}^{\pm}\left(  \omega \right)  =\pm \omega_{\vec{q}}+\frac{\left(  \vec
{k}+\vec{q}\right)  ^{2}}{2\left(  1+M\right)  }+\left(  n-n^{\prime}\right)
\Omega-\omega.
\]
As shown in Ref. \cite{PhysRevB.28.6051} this expression has a very natural
interpretation: every term in the double sum corresponds to a well-defined
physical process. The ($n$,$n^{\prime}$)th term in the summation represents a
scattering process of the polaron where in the initial state the polaron is in
the $n^{\prime}$th internal Franck-Condon state of the Feynman polaron model
system and is then scattered to the $n$th internal Franck-Condon state with
the absorption of an energy $\hbar \omega$ and a momentum $\vec{k}$ through the
Bragg scattering and with the emission (or absorption) of a Bogoliubov
excitation with energy $\hbar \omega_{\vec{q}}$.

\section{Sum rule}

As a consistency check of our results we can apply the f-sum rule to our
results, this can be written as:%
\begin{equation}
\int_{0}^{\infty}d\omega \omega \operatorname{Im}\left[  \chi \left(  \omega
,\vec{k}\right)  \right]  =N\frac{\pi}{2}\frac{k^{2}}{m_{I}}.
\end{equation}
At small $\omega$ and as temperature tends to zero a delta peak appears in the
integrand for which, in numerical calculations, we have to add the
contribution separately. This feature was found for the solid state
Fr\"{o}hlich polaron, where the contribution of the delta peak can be used to
determine the effective mass of the polaron \cite{PhysRevB.15.1212}. The delta
peak can be revealed if we look at the $\omega \rightarrow0$ limits of the
imaginary and real part of expression (\ref{SelfEnergie}) for the memory
function:%
\begin{align}
\lim_{\omega \rightarrow0}\operatorname{Im}\left[  \Sigma \left(  \omega,\vec
{k}\right)  \right]   &  =0;\\
\lim_{\omega \rightarrow0}\frac{\operatorname{Re}\left[  \Sigma \left(
\omega,\vec{k}\right)  \right]  }{\omega^{2}}  &  =R\left(  \alpha,\vec
{k}\right)  ,
\end{align}
with:%
\begin{align}
R\left(  \alpha,\vec{k}\right)   &  =\frac{\omega_{0}^{2}}{N}\frac{\alpha
}{4\pi}\left(  \frac{m_{B}+1}{m_{B}}\right)  ^{2}\int_{0}^{\infty}%
d\widetilde{q}\widetilde{q}^{2}\frac{q}{\sqrt{q+2}}q^{2}\int_{0}^{\infty
}dt\frac{t^{2}}{2}\nonumber \\
&  \times \operatorname{Im}\left[  \frac{\left[  e^{i\omega_{\vec{q}}t}%
+2\cos \left(  \omega_{\vec{q}}t\right)  n_{\vec{q}}\right]  }{\left[
2kqD\left(  t\right)  \right]  ^{3}}\left(  \exp \left[  -\left(  k-q\right)
^{2}D\left(  t\right)  \right]  \left \{  \left[  1-2kqD\left(  t\right)
\right]  ^{2}+1\right \}  \right.  \right. \nonumber \\
&  \left.  \left.  -\exp \left[  -\left(  k+q\right)  ^{2}D\left(  t\right)
\right]  \left \{  \left[  1+2kqD\left(  t\right)  \right]  ^{2}+1\right \}
\right)  \right]  .
\end{align}
If we now look at the $\omega \rightarrow0$ limit of the density response
function we find:%
\begin{equation}
\lim_{\omega \rightarrow0}\operatorname{Im}\left[  \chi \left(  \omega,\vec
{k}\right)  \right]  =\frac{k^{2}N}{m_{I}\left(  1-R\left(  \alpha \right)
\right)  }\pi \delta \left(  \omega^{2}\right)  ,
\end{equation}
where we used the following representation for the delta function:%
\begin{equation}
\delta \left(  x\right)  =\lim_{n\rightarrow0}\frac{1}{n\pi}\frac{1}{1+\left(
x/n\right)  ^{2}}.
\end{equation}
So at temperature zero the sum rule becomes:%
\begin{equation}
\frac{\pi}{2}\frac{k^{2}N}{m_{I}\left(  1-R\left(  \alpha,\vec{k}\right)
\right)  }+\int_{\varepsilon}^{\infty}d\omega \omega \operatorname{Im}\left[
\chi \left(  \omega,\vec{k}\right)  \right]  =N\frac{\pi}{2}\frac{k^{2}}{m_{I}%
}, \label{sumrule}%
\end{equation}
where $\varepsilon$ is a positive infinitesimal. Care has to be taken at
finite temperature since the delta peak will then broaden and start to overlap
with other contributions of the spectrum. At low enough temperatures we will
see that our results agree well with (\ref{sumrule}).

\section{Extended memory function formalism\label{extended}}

It was noticed in Ref. \cite{Cataudella} that the Feynman polaron model system
does not satisfy the sum rule for the density-density correlation. This
limitation can be removed by introducing a finite lifetime $\tau$ for the
eigenstates of the Feynman model system which mimics scattering events with
the bosonic bath, which can be done by replacing $\exp \left(  i\Omega
t\right)  $ by $\left(  1+it/\tau \right)  ^{\Omega \tau}$ in (\ref{Dfun}). This
technique is called the extended memory function formalism and was applied in
Ref. \cite{PhysRevLett.96.136405} for the solid state Fr\"{o}hlich polaron. It
was shown that at intermediate coupling this gives a better agreement between
the theory and the Diagrammatic Quantum Monte-Carlo numerical techniques. This
extension also solves the problem of the inconsistency of the linewidths at
strong coupling with the uncertainty relation for the solid state Fr\"{o}hlich
polaron, which was first mentioned in Ref. \cite{PhysRevB.5.2367}. It seems
that the second representation for the imaginary part of the memory function
(\ref{TweedeRepr}) is not suitable for the introduction of a lifetime. To
obtain a qualitatively picture of the dependence of the results on the
lifetime we can work with the real and imaginary part of the memory function
as expressed in equation (\ref{SelfEnergie}).

\section{Results and discussion}

The results we obtain in this section are for a system consisting of a sodium
condensate with a single lithium impurity, this means that we use for the
bosonic mass: $m_{B}=3.8221$, which is in polaronic units as are all the
results in this section.

\begin{figure}
[ptb]
\begin{center}
\includegraphics[
height=3.9166cm,
width=8.7052cm
]%
{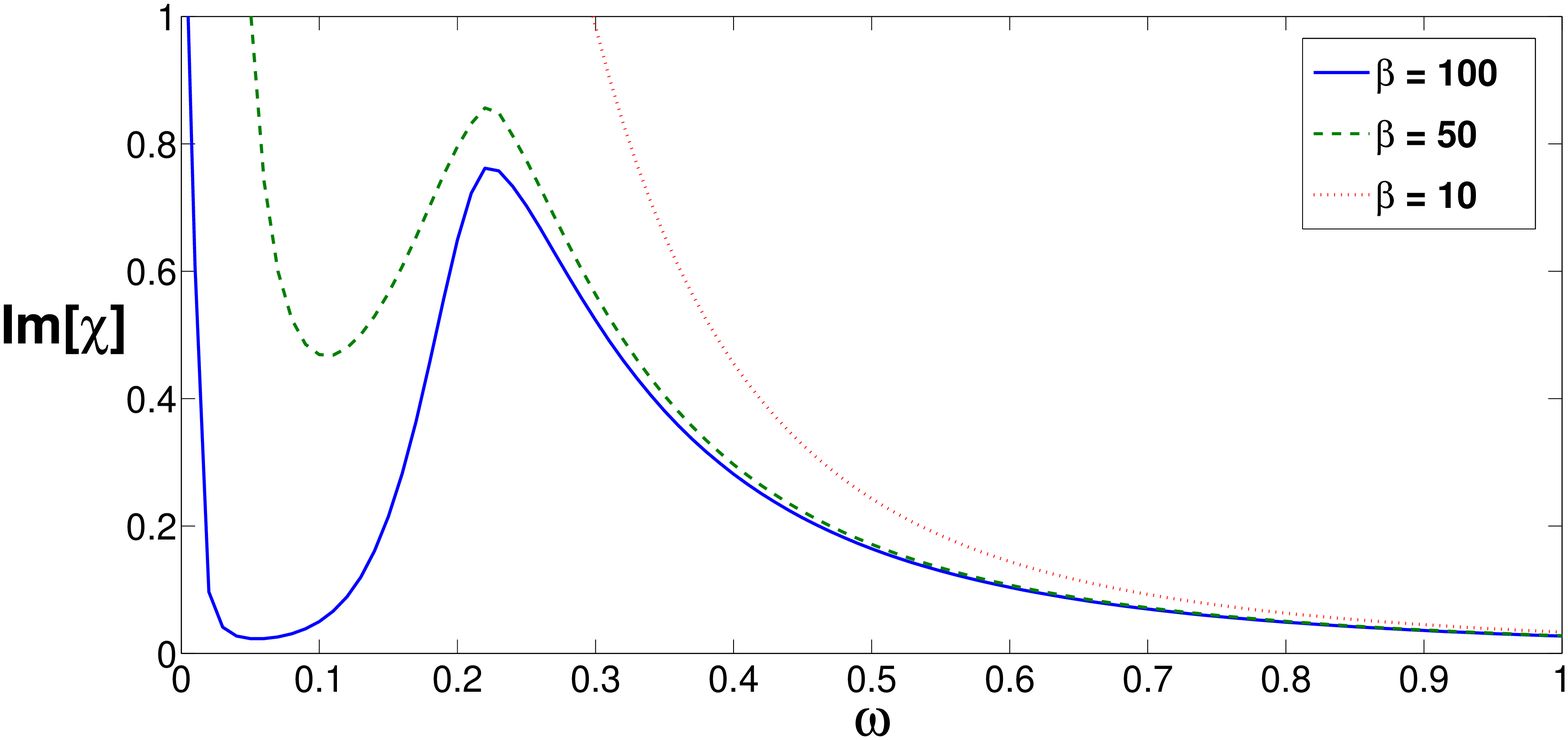}%
\caption{The Bragg response of the polaron system consisting of an impurity in
a BEC polaron in the weak coupling regime ($\alpha=0.1$) with momentum
exchange $k=1$ for different temperatures.}%
\label{fig: zwakbafh}%
\end{center}
\end{figure}

We begin by looking at the weak coupling regime. In figure \ref{fig: zwakbafh}
the Bragg response (\ref{repons}) is presented as a function of the
transferred energy $\hbar \omega$ for different temperatures and for a momentum
exchange $k=1$. At low temperature ($\beta=100$) we clearly see a peak that
represents the weak coupling scattering process and can be understood as the
emission of Bogoliubov excitations. Also the contribution of the temperature
broadened delta peak at low $\omega$ is seen. This is the anomalous Drude peak
(see Ref. \cite{Devreese1998309}). If we look at higher temperatures the zero
temperature delta peak broadens and there is a larger overlap with the weak
coupling scattering peak. At $\beta=10$ this peak has become indistinguishable
from the anomalous Drude peak. For $\beta=100$ a distinction between the two
contributions can still be made and the f-sum rule can be applied, as will be
done below.%

\begin{figure}
[ptb]
\begin{center}
\includegraphics[
trim=0.000000in 0.000000in 0.001624in 0.000000in,
height=3.9144cm,
width=8.703cm
]%
{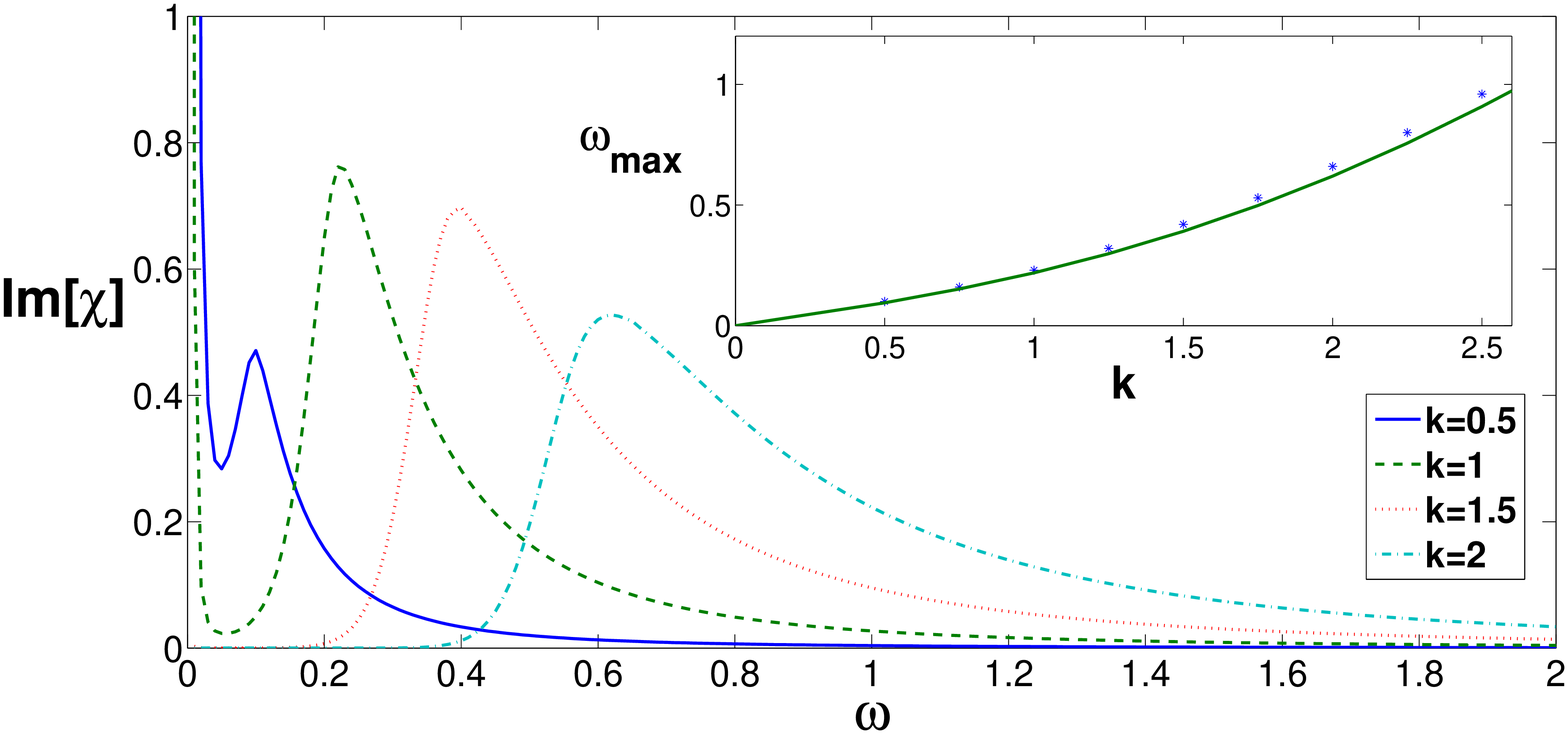}%
\caption{The Bragg response of the polaron system consisting of an impurity in
a BEC in the weak coupling regime ($\alpha=0.1$) at temperature $\beta=100$
for different momentum exchanges. The inset shows the frequency og the maximum
as a function of the exchanged momentum $k$. The curve gives a least square
fit of the Bogoliubov spectrum (\ref{FitBogSpect}), with fitting parameter
$m=3.9534$.}%
\label{fig: weakqafh}%
\end{center}
\end{figure}

The dependence of the spectrum on the exchanged momentum $k$ is presented in
figure \ref{fig: weakqafh} for $\alpha=0.1$ and $\beta=100$. For larger
momentum exchange the scattering peak is shifted to higher frequencies and a
broadening is observed. The inset of figure \ref{fig: weakqafh} shows the
frequencies $\omega_{\max}$ at which the maximum of the peak occurs as a
function of the exchanged momentum $k$ together with a least square fit to the
Bogoliubov spectrum:%
\begin{equation}
\omega=\frac{k}{2m}\sqrt{k^{2}+2},\label{FitBogSpect}%
\end{equation}
where $m$ is determined as a fitting parameter to be: $m=3.9534$; which is in
good agreement with the bosonic mass of the condensate ($m_{B}=3.8221$). This
shift according to the Bogoliubov dispersion is plausible since the peak
corresponds to the emission of Bogoliubov excitations.%

\begin{figure}
[ptb]
\begin{center}
\includegraphics[
height=3.9144cm,
width=8.703cm
]%
{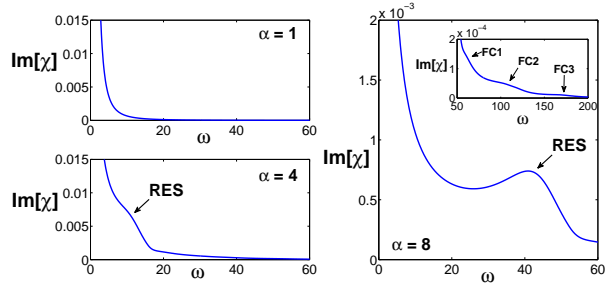}%
\caption{The Bragg response of the polaron system consisting of an impurity in
a BEC at different values for the coupling parameter $\alpha$ at temperature
$\beta=100$ and a momentum exchange $k=1$. In the strong coupling regime the
imprint of the Relaxed Excited State resonance appears. The inset shows the
first three Franck-Condon peaks at $\alpha=8$.}%
\label{fig: RES}%
\end{center}
\end{figure}

In figure \ref{fig: RES} the high frequency tail of the Bragg spectrum is
shown for different coupling strengths $\alpha$ at $\beta=100$ and $k=1$. In
the strong coupling regime a resonance is seen which is absent in the weak
coupling regime. This feature is well-known from the solid state Fr\"{o}hlich
polaron and corresponds to a transition to the Relaxed Excited State (RES); it
was first proposed in \cite{PhysRevLett.22.94}. This resonance appears at a
frequency $\omega_{RES}$ such that $\omega_{RES}^{2}=\operatorname{Re}\left[
\Sigma \left(  \omega_{RES},\vec{k}\right)  \right]  $ with the supplementary
condition $\operatorname{Im}\left[  \Sigma \left(  \omega_{RES},\vec{k}\right)
\right]  \ll1$. It is clear from (\ref{repons}) that these conditions cause a
peak in the spectrum. This resonance corresponds to a transition from the
polaron ground state to an excited state in the polaronic self-trapping
potential which has been relaxed consistent with the new excited wave function
of the impurity. The coupling strength where the relaxed excited state appears
in the Bragg spectrum is slightly below $\alpha=4$. This is in agreement with
the prediction in \cite{PhysRevB.80.184504} that for a BEC-impurity the
transition between the weak and the strong coupling regime occurs around
$\alpha \simeq3$ for $\beta \rightarrow \infty$. In the strong coupling regime
other peaks are present which are indicated for $\alpha=8$ in the inset of
figure \ref{fig: RES}. These are the Franck-Condon (FC) peaks and represent a
transition to the RES together with the emission of Bogoliubov excitations.
They only appear in the strong coupling regime which was also observed in the
case of the acoustic polaron \cite{PhysRevB.35.3745}.

\begin{figure}
[ptb]
\begin{center}
\includegraphics[
height=3.9166cm,
width=8.7052cm
]%
{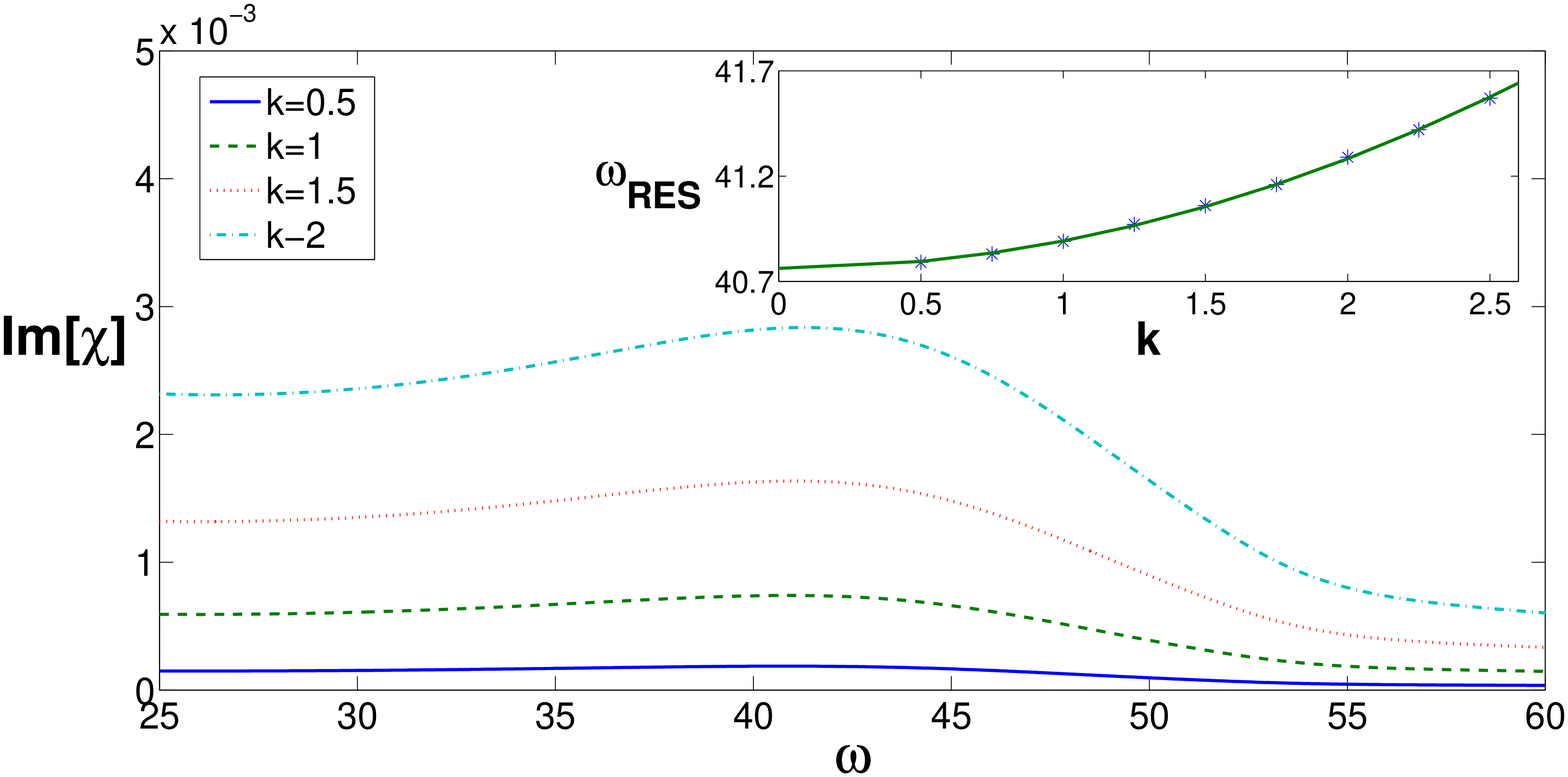}%
\caption{The Realaxed Excited State peak in the Bragg response of the polaron
system consisting of an impurity in a BEC in the strong coupling regime
($\alpha=8$) at temperature $\beta=100$ for different momentum exchanges. The
inset shows the frequency of the maximum as a function of the exchanged
momentum $k$ at $\beta=100$. The curve is a least square fit to the quadratic
dispersion (\ref{QuadrDisp}), with fitting parameters $\nu=40.76$ and
$m^{\ast}=3.84$.}%
\label{fig: strongqdep}%
\end{center}
\end{figure}

The dependence of the RES peak on the exchanged momentum is depicted in figure
\ref{fig: strongqdep}. The inset shows the frequency of the maximum of the RES
peaks as a function of the exchanged momentum $q$ together with a least square
fit to a quadratic dispersion:%
\begin{equation}
\omega=\nu+\frac{k^{2}}{2m^{\ast}},\label{QuadrDisp}%
\end{equation}
where $\nu$ and $m^{\ast}$ are the fitting parameters. This suggests that the
RES \ is characterized by a transition frequency $\nu$ and an effective mass
$m^{\ast}$.%

As a consistency test it was checked whether the above results satisfy the
f-sum rule (\ref{sumrule}). Filling out the expression for the imaginary part
of the density response function (\ref{repons}) and dividing by the common
factor the f-sum rule takes the form:%
\begin{equation}
\frac{\pi}{2\left(  1-R\left(  \alpha,\vec{k}\right)  \right)  }%
-\int_{\varepsilon}^{\infty}d\omega \omega \frac{\operatorname{Im}\left[
\Sigma \left(  \omega,\vec{k}\right)  \right]  }{\left(  \omega^{2}%
-\operatorname{Re}\left[  \Sigma \left(  \omega,\vec{k}\right)  \right]
\right)  ^{2}+\left(  \operatorname{Im}\left[  \Sigma \left(  \omega,\vec
{k}\right)  \right]  \right)  ^{2}}=\frac{\pi}{2}. \label{sunrule2}%
\end{equation}
It is impossible to integrate numerically to infinity and for this reason a
cut-off, $\omega_{c}=500$, is used. A calculation of the left hand side of
(\ref{sunrule2}) results in the values in table \ref{Table: SumRule} for
different values of $\alpha$ and $k$ at $\beta=100$. These results should be
compared with the rigourous value $\pi/2=1.5708...$ which gives a good
agreement with small deviations.%

\begin{table}[tbp] \centering
\begin{tabular}
[c]{l|llll}%
$k$ & $\alpha=$ & $1$ & $4$ & $8$\\ \hline
$1$ &  & $1.5442$ & $1.6845$ & $1.5532$\\
$3$ &  & $1.5482$ & $1.6419$ & $1.5355$\\
$5$ &  & $1.5506$ & $1.5555$ & $1.4979$%
\end{tabular}
\caption{Results of a numerical calculation of the left hand side of expression (\ref{sunrule2}) with a cut-off $\omega_c = 500$ introduced for the $\omega$-integral. }\label{Table: SumRule}%
\end{table}%

\begin{figure}
[ptb]
\begin{center}
\includegraphics[
height=3.9166cm,
width=8.7052cm
]%
{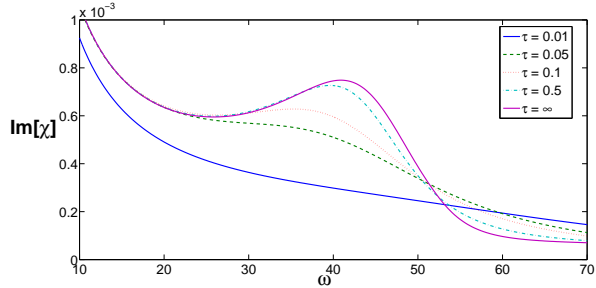}%
\caption{The Relaxed Excited State peak in the Bragg response of the polaron
system consisting of an impurity in a BEC in the strong coupling regime
($\alpha=8$) at temperature $\beta=100$ with different lifetimes $\tau$. }%
\label{fig: lifetimes}%
\end{center}
\end{figure}

We now give a qualitative analysis of the dependence of the results on the
lifetime $\tau$, which was introduced in section \ref{extended} within the
extended memory function formalism. In figure \ref{fig: lifetimes} the Relaxed
Excited State peak is presented at different values for $\tau$ in the strong
coupling regime ($\alpha=8$). It is observed that the inclusion of a lifetime
parameter of the order of the polaronic time unit results in a broadening of
the peak, where a smaller lifetime corresponds to a broader peak. When a
lifetime of the order of a hundredth of the polaronic time unit is introduced
the peak can not be distinguished any more.%

\section{Conclusions}

We have derived a general formula for the density response function as a
function of the transferred energy and momentum with the Mori-Zwanzig
projection operator formalism. This provides a general result that can be used
for Bragg spectroscopy but also for other probes that exhibit an arbitrary
energy and momentum exchange as for example neutron scattering where the
output is also determined by the density response function
\cite{PhysRev.95.249}. This is applied to the Fr\"{o}hlich polaron Hamiltonian
for which the well-known results from Ref. \cite{PhysRevB.5.2367} for the
optical absorption are found. We then extend the analysis to Bragg scattering
of impurity polarons in a Bose condensed gas, where the Bogoliubov excitations
play the role of the phonons in the polaron formation. The f-sum rule is checked.

To analyze the results we introduced the specific system of a lithium impurity
in a sodium condensate and calculated the spectra in the different coupling
regimes and for different momentum exchanges and temperatures. It is seen that
these spectra possess similar features as also found in the optical absorption
of the solid state Fr\"{o}hlich polaron. Furthermore it was shown that the
weak coupling scattering peaks follow the Bogoliubov spectrum as a function of
the exchanged momentum. In the strong coupling regime the Relaxed Excited
State emerges, and we derive the transition frequency and the effective mass
associated with the Relaxed Excited State. This is of importance for the
comparison with Diagrammatic Quantum Monte-Carlo numerical techniques, which
in the case of the optical absorption of the solid state Fr\"{o}hlich polaron
has led to new results concerning the linewidth and oscillator strength of the
Relaxed Excited State and Franck-Condon transitions. The Franck-Condon peaks
were also observed.

Our results were tested using the f-sum rule which resulted in a good
agreement with small deviations.

The influence of the introduction of a lifetime within the extended memory
function formalism was also qualitatively investigated and it was shown that
this results in a broadening of the RES peak.

\begin{acknowledgments}
This work was supported by FWO-V under Project Nos. G.0180.09N, G.0115.06,
G.0356.06, G.0370.09N, and the WOG Belgium under Project No. WO.033.09N. J.T.
gratefully acknowledges support of the Special Research Fund of the University
of Antwerp, BOF NOI UA 2004. M.O. acknowledges financial support by the
ExtreMe Matter Institute EMMI in the framework of the Helmholtz Alliance under
Grant No. HA216/EMMI. W. C. acknowledges financial support from the BOF-UA. 
\end{acknowledgments}

\bibliographystyle{Science}
\bibliography{respons}

\end{document}